# Spin and Orbital Effects on Asymmetric Exchange Interaction in Polar Magnets: M(IO$_3$)$_2$ (M = Cu, Mn)

Ebube E. Oyeka,[1] Michał J. Winiarski,[2] Maurice Sorolla II,[3] Keith M. Taddei,[4] Allen Scheie,[4] and Thao T. Tran*,[1]

[1]Department of Chemistry, Clemson University, Clemson, SC 29634, USA

[2]Faculty of Applied Physics and Mathematics and Advanced Materials Center, Gdansk University of Technology, ul. Narutowicza 11/12, 80-233 Gdansk, Poland

[3]Department of Chemical Engineering, University of the Philippines Diliman, Quezon City 1101, Philippines

[4]Neutron Scattering Division, Oak Ridge National Laboratory, 9500 Spallation Dr, Oak Ridge, TN 37830, USA

**ABSTRACT:** Magnetic polar materials feature an astonishing range of physical properties, such as magnetoelectric coupling, chiral spin textures and related new spin topology physics. This is primarily attributable to their lack of space inversion symmetry in conjunction with unpaired electrons, potentially facilitating an asymmetric Dzyaloshinskii-Moriya (DM) exchange interaction supported by spin-orbital and electron-lattice coupling. However, engineering the appropriate ensemble of coupled degrees of freedom necessary for enhanced DM exchange has remained elusive for polar magnets. Here we study how spin and orbital components influence the capability of promoting the magnetic interaction by studying the two magnetic polar materials, $\alpha$-Cu(IO$_3$)$_2$ ($^2$D) and Mn(IO$_3$)$_2$ ($^6$S) and connecting their electronic and magnetic properties with their structures. The chemically controlled low-temperature synthesis of these complexes resulted in pure polycrystalline samples, providing a viable pathway to prepare bulk forms of transition-metal iodates. Rietveld refinements of the powder synchrotron X-ray diffraction data reveal that these materials exhibit different crystal structures but crystallize in the same polar and chiral $P2_1$ space group, giving rise to an electric polarization along the $b$-axis direction. The presence and absence of an evident phase transition to a possible topologically distinct state observed in $\alpha$-Cu(IO$_3$)$_2$ and Mn(IO$_3$)$_2$, respectively, implies the important role of spin orbit coupling. Neutron diffraction experiments reveal helpful insights into the magnetic ground state of these materials. While the long-wavelength incommensurability of $\alpha$-Cu(IO$_3$)$_2$ is in harmony with sizable asymmetric DM interaction and low dimensionality of electronic structure, the commensurate stripe AFM ground state of Mn(IO$_3$)$_2$ is attributed to negligible DM exchange and isotropic orbital overlapping. The work demonstrates connections between combined spin and orbital effects, magnetic coupling dimensionality and DM exchange, providing a worthwhile approach for tuning asymmetric interaction which promotes evolution of topologically distinct spin phases.

## INTRODUCTION

Polar magnets display an intriguing spectrum of novel physical properties, from multiferroics to vortex-like spin states, and associated nontrivial topology spin physics.[1-13] Topologically distinct spin textures, such as skyrmions, featuring a nanometric-sized particle nature are anticipated to play an important role in several areas, from new knowledge in electronic and magnetic states of matter, to unconventional computing and high-density low-power magnetic memory applications.[14-19] Underpinning much of this behavior is the ability of magnetic polar compounds, in which spatial inversion symmetry is broken, to stabilize Dzyaloshinskii-Moriya (DM) exchange interaction supported by strong coupling of spin, orbital and lattice degrees of freedom.[11, 15, 17, 20, 21]

Notably, correlations between these quantum effects go hand-in-hand with the relativistic spin-orbit coupling (SOC),[22] which has a tendency to twist the mutual spin alignment through the DM interaction.[17, 23] Enhanced DM exchange in the presence of the Heisenberg interaction in polar magnets promotes stabilizing a winding spin lattice in a certain range of applied magnetic fields and temperatures. Manifestations of SOC also substantially impact the characteristic quantities, such as transition temperature, critical field, spin topology, magnetic modulation wavelength and electronic structure, in magnetic phase diagrams of skyrmion host materials. MnSi, a B20-structure chiral silicide metal, undergoes a transition to a Bloch skyrmion phase at $T$ = 30 K and $H_c$ = 0.2 T with the magnetic modulation period $\lambda$ = 18 nm.[24, 25] Interestingly, partial or complete chemical substitution of Si atoms in MnSi with Ge atoms in MnGe and MnSi$_{1-x}$Ge$_x$, which can induce larger SOC, significantly changes the magnetic behavior including an enhanced magnetic ordering temperature ($T_N$ = 170 K) and the formation of 3D hedgehog spin lattice states with very small sizes ($\lambda$ = 3 - 6 nm).[26-28] In addition to these spin and orbital effects, it has been suggested that lattice symmetry plays a role in initiating topologically nontrivial spin configurations. Take Cu$_2$OSeO$_3$ as an example: the magnetic Cu$^{2+}$ cations with orbital effects residing on the 3-fold rotation axis has been thought to enable stabilization of its skyrmion lattice.[2] In Fe(IO$_3$)$_3$, a combination of stereoactive lone-pair electron effects and $C_3$ magnetic cations symmetry was proposed to work in favor of improved DM exchange.[29] Yet compounds with the point group of $D_{2d}$ and $C_{4v}$, in which 3-fold symmetry is absent, display unique characters concerning asymmetric DM interactions, thus hosting antiskyrmion and Neel skyrmion as exemplified in Mn$_{1.4}$Pt$_{0.9}$Pd$_{0.1}$Sn and VOSe$_2$O$_5$, respectively.[11, 30]

However, due to the scarcity of systematic studies of polar magnets featuring enhanced or hindered DM exchange, synthetic tunability of spin properties, deterministic engineering of orbital contribution and demonstration of how the coupled spin and orbital degrees of freedom drive the potential emergence of vortex-like spin textures still remain outstanding goals. To help address such fundamental questions, our study focused on how spin and orbital contribution can cause increased or reduced magnetic interactions in two polar magnets: α-Cu(IO$_3$)$_2$ (Cu$^{2+}$, d$^9$, $^2$D) and Mn(IO$_3$)$_2$ (Mn$^{2+}$, d$^5$, $^6$S).

## EXPERIMENTAL SECTION

**Reagents.** CuCl$_2$ (Alfa Aesar, 99.9 %), MnCl$_2$ (Alfa Aesar, 99.9 %), ZnCl$_2$ (Alfa Aesar, 99.95 %), HIO$_3$ (Alfa Aesar, 99.5 %), HNO$_3$ (BDH, 67 %) were used as starting materials.

**Synthesis. α-Cu(IO$_3$)$_2$.** Polycrystalline α-Cu(IO$_3$)$_2$ was synthesized by heating Cu(IO$_3$)$_2$·2/3H$_2$O at 300 °C for 8 hours (Figure S1). For Cu(IO$_3$)$_2$·2/3H$_2$O, anhydrous CuCl$_2$ (2.5 mmol) was dissolved in HNO$_3$ (1M, 25 mL) and then an aqueous solution of HIO$_3$ (5 mmol, 10 mL) was added. The mixture was stirred at 80 °C for 1 h. Greenish polycrystalline Cu(IO$_3$)$_2$·2/3H$_2$O was isolated from the mother-liquor after 24 h by filtration, washed with deionized water and dried in air (yield of α-Cu(IO$_3$)$_2$, 92 %; based on Cu).

**Mn(IO$_3$)$_2$.** Polycrystalline Mn(IO$_3$)$_2$ was synthesized by the reaction of MnCl$_2$ and HIO$_3$ at 80 °C and ambient pressure. MnCl$_2$ (2.5 mmol) was dissolved in HNO$_3$ (1 M, 25 mL). An aqueous solution of HIO$_3$ (5 mmol, 10 mL) was added and the mixture was stirred at 80 °C for 1 h. A pale pink solid was isolated from the reaction mixture by filtration and washed with deionized water. The solid was dried in air for 1 week to give anhydrous polycrystalline Mn(IO$_3$)$_2$ (yield of Mn(IO$_3$)$_2$, 86 %; based on Mn) (Figure S2).

**Zn(IO$_3$)$_2$.** ZnCl$_2$ (2.5 mmol) was dissolved in HNO$_3$ (1 M, 25 mL). An aqueous solution of HIO$_3$ (5 mmol, 10 mL) was added and the mixture was stirred at 80 °C for 1 h to give a colorless solution. Colorless solid of polycrystalline Zn(IO$_3$)$_2$ precipitated from the solution after 3 days and was collected by filtration and dried in air (yield of Zn(IO$_3$)$_2$, 65 %; based on Zn) (Figure S3).

**Synchrotron X-ray Diffraction.** Synchrotron XRD patterns of α-Cu(IO$_3$)$_2$ and Mn(IO$_3$)$_2$ were collected using the 11-BM beamline at Advanced Photon Source, Argonne National Laboratory. Data were collected at $T$ = 295 K and $\lambda$ = 0.45789 Å. No impurities were observed. Rietveld refinement of the Synchrotron XRD patterns was performed using TOPAS Academic V6. VESTA software was used for crystal structure visualization.[31]

**Powder X-ray Diffraction.** Powder X-ray diffraction (PXRD) measurements on Zn(IO$_3$)$_2$ and Cu(IO$_3$)$_2$·2/3H$_2$O were performed using Rigaku Ultima IV diffractometer equipped with Cu $K_\alpha$ radiation ($\lambda$ = 1.5406 Å). Data were collected in the $Q$ range of 0.36 Å$^{-1}$ – 5.78 Å$^{-1}$ at 0.0142 Å$^{-1}$/min. Rietveld refinement of XRD pattern was performed using TOPAS Academic V6 (Figure S4-S5).

**Thermal Analysis.** Thermogravimetric analysis (TGA) and differential scanning calorimetry (DSC) measurements were performed using a TA SDT Q600 instrument. Approximately 10 mg of Cu(IO$_3$)$_2$·2/3H$_2$O, Mn(IO$_3$)$_2$ or Zn(IO$_3$)$_2$ was placed separately in an alumina crucible and heated at a rate of 20 °C/min from room temperature to 1000 °C under flowing nitrogen (flow rate: 100 mL/min). [Figure S1-S3]

**Infrared Spectroscopy.** Attenuated total reflection Fourier transform infrared (ATR-FTIR) spectra for α-Cu(IO$_3$)$_2$, Mn(IO$_3$)$_2$, and Zn(IO$_3$)$_2$ were collected separately using a Shimadzu IR Affinity-1S in 400 to 4000 cm$^{-1}$ frequency range (Figure S6-S7).

**Magnetization and Specific Heat.** DC Magnetization measurements on α-Cu(IO$_3$)$_2$ and Mn(IO$_3$)$_2$ powders were performed with the Vibrating sample magnetometer (VSM) option of Quantum Design Physical Properties Measurement System (PPMS) using the standard polyethylene straw sample holders. Data were collected under the applied magnetic field of $\mu_0H$ = 5 T from $T$ = 2 K – 300 K and $\mu_0H$ = 0.05 T from $T$ = 2 K – 25 K. Magnetic susceptibility was approximated as magnetization divided by the applied magnetic field: $\chi \approx M/H$. High resolution DC magnetization data of α-Cu(IO$_3$)$_2$ and Mn(IO$_3$)$_2$ were collected using the VSM option of Quantum Design Physical Properties Measurement System (PPMS) around the magnetic transition temperature from $\mu_0H$ = 0.4 T – 4 T (Figure S8-S9). Magnetic entropy calculations were performed using the dedicated script by Bocarsly et al.[32] Heat capacity was measured using the PPMS, employing the semiadiabatic pulse technique from $T$ = 2 K – 300 K (Figure S10).

**Neutron Diffraction.** Neutron diffraction data for α-Cu(IO$_3$)$_2$ and Mn(IO$_3$)$_2$ were measured using the HB2A diffractometer at ORNL's HFIR reactor.[33] 5 g of each powder were loaded into aluminum sample cans under 1 atm helium, and cooled to a base temperature of 1.6 K. We measured the diffraction patterns with neutrons of incident wavelength 2.41 Å in two separate experiments (both diffraction patterns and the Cu order parameter curve were measured in the first experiment, and the Mn order parameter curve was measured in the second). For each compound, we measured the diffraction pattern above and below the Neel temperature, which showed the onset of magnetic peaks at low temperatures associated with long range magnetic order. To verify that these did indeed appear at the critical temperature, we measured the Bragg intensity as a function of temperature through the phase transitions of several of the low temperature peaks.

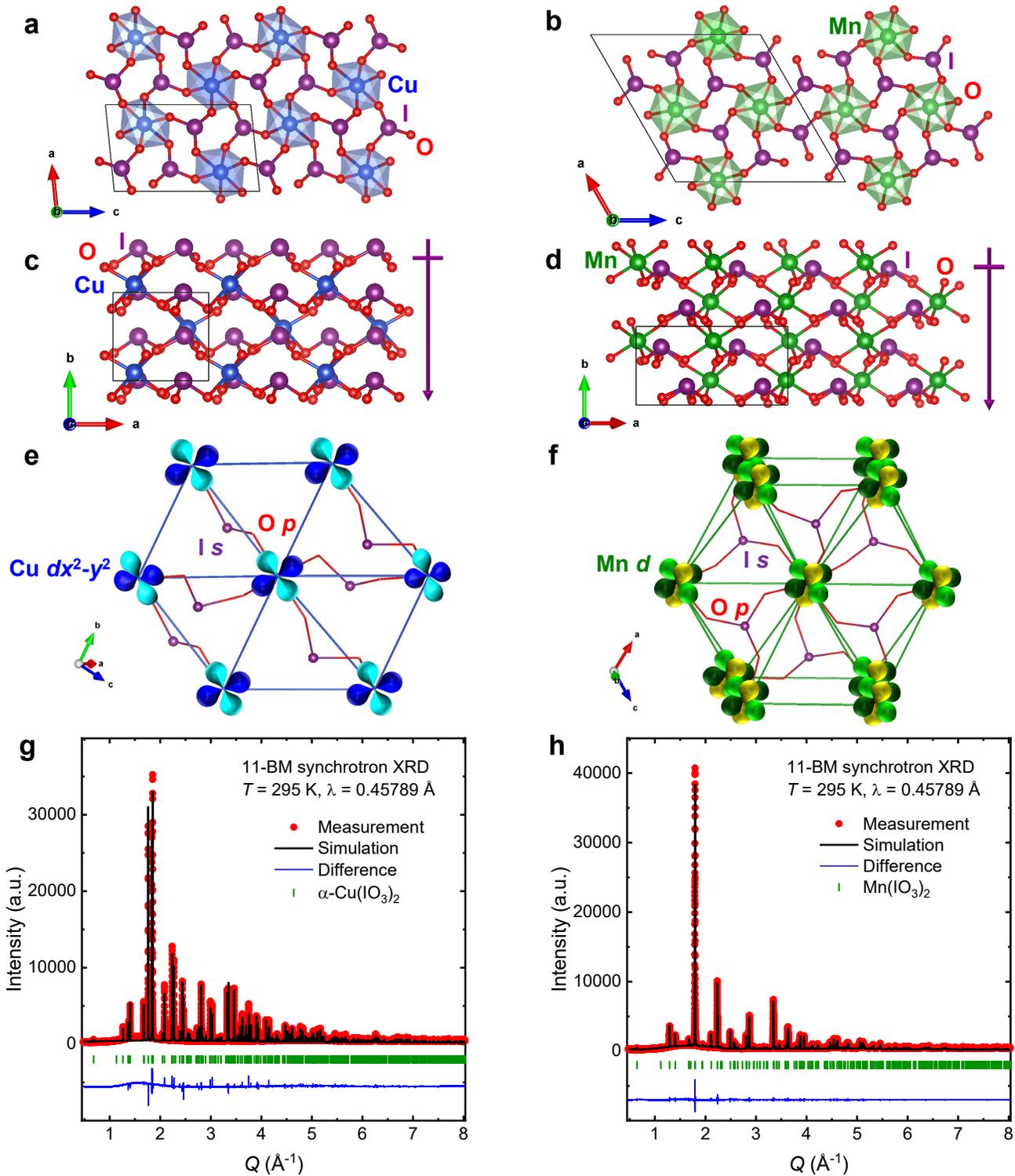

**Figure 1.** (a-b) Crystal structure of α-Cu(IO$_3$)$_2$ and Mn(IO$_3$)$_2$ respectively showing MO$_6$ sub-unit in a slightly distorted octahedral geometry and the trigonal pyramidal (IO$_3$)$^-$. (c-d) Crystal structure of α-Cu(IO$_3$)$_2$ and Mn(IO$_3$)$_2$ respectively showing MO$_6$ octahedra and (IO$_3$)$^-$ trigonal pyramid aligned along the polar *b* axis. (e-f) Atomic connectivity and orbital overlap in α-Cu(IO$_3$)$_2$ and Mn(IO$_3$)$_2$ respectively. (g-h) Rietveld fit (black line) of high resolution synchrotron XRD patterns (red points) of polycrystalline α-Cu(IO$_3$)$_2$ and Mn(IO$_3$)$_2$ respectively. The blue lines denote the difference between the measured data and the simulation.

**DFT Calculations.** Spin-polarized electronic structure calculations were performed using full-potential linearized augmented plane wave method as implemented in the WIEN2k code.[34] The exchange and correlation energies were treated within the density functional theory (DFT) using the Perdew–Burke–Ernzerhof generalized gradient approximation.[35] The muffin-tin radius values 1.87, 2.04, 1.73 and 1.49 au were used for Cu, Mn, I and O, respectively. The self-consistencies were carried out using 1000 $k$-points (10 × 10 × 10 mesh) in the irreducible Brillouin zone.

## RESULTS AND DISCUSSION

**Synthesis.** α-Cu(IO$_3$)$_2$ was obtained by annealing Cu(IO$_3$)$_2$·2/3H$_2$O at 300 °C, while Mn(IO$_3$)$_2$ was isolated in the anhydrous form from the reaction mixture of MnCl$_2$ and HIO$_3$. Polycrystalline α-Cu(IO$_3$)$_2$ and Mn(IO$_3$)$_2$ are brown and light pink, respectively. The brown color of α-Cu(IO$_3$)$_2$ was distinct from the green color of Cu(IO$_3$)$_2$·2/3H$_2$O. The synthesis of anhydrous polycrystalline α-Cu(IO$_3$)$_2$ was based on the thermal behavior of its precursor Cu(IO$_3$)$_2$·2/3H$_2$O. To determine the annealing temperature for α-Cu(IO$_3$)$_2$, the thermal properties of the hydrated precursor were characterized by thermogravimetric analysis (TGA) and differential scanning calorimetry (DSC). From the TGA/DSC analysis of Cu(IO$_3$)$_2$·2/3H$_2$O (Figure S1), the endothermic peak at ~230 °C corresponds to the loss of 2/3H$_2$O. The experimental mass loss of 2.9 % is in excellent agreement with the calculated value of 2.9 %. From the TGA/DSC analysis of Cu(IO$_3$)$_2$·2/3H$_2$O, it was inferred that α-Cu(IO$_3$)$_2$ forms as a stable phase at ~300 °C and this was confirmed by X-ray diffraction. α-Cu(IO$_3$)$_2$ and Mn(IO$_3$)$_2$ possess stability to air and moderate temperature. The decomposition temperatures were determined to be 492 °C and 448 °C for α-Cu(IO$_3$)$_2$ and Mn(IO$_3$)$_2$, respectively, accompanied by strong endothermic transitions in the DSC curves. This is attributed to the loss of I$_2$ and 2.5O$_2$. The experimental mass loss is in good agreement with the calculated mass loss (α-Cu(IO$_3$)$_2$: Exp. 81.5 %, Cal. 81.3 %; Mn(IO$_3$)$_2$: Exp. 82.7 %, Cal. 82.5 %).

**Crystal Structure.** In order to evaluate the quality of α-Cu(IO$_3$)$_2$ and Mn(IO$_3$)$_2$ bulk samples prepared from the aforementioned synthetic methodologies, their synchrotron XRD data were collected using the 11-BM beamline equipped with high reciprocal space resolution and counting statistics at Advanced Photon Source, Argonne National Laboratory. While the crystal structures of α-Cu(IO$_3$)$_2$ and Mn(IO$_3$)$_2$ were previously reported,[36, 37] it is helpful to confirm their lattice symmetry, especially with possible subtle distortions driven by spin-orbit coupling that we might expect to observe in these polar magnets. α-Cu(IO$_3$)$_2$ and Mn(IO$_3$)$_2$ crystal structures obtained from Rietveld refinements of X-ray synchrotron powder diffraction data are in excellent agreement with those reported.[36, 37] It is worth noticing that α-Cu(IO$_3$)$_2$ and Mn(IO$_3$)$_2$ crystallize in the same monoclinic non-centrosymmetric polar $P2_1$ space group (Figure 1), nevertheless, their crystal structures are different.

α-Cu(IO$_3$)$_2$ structure consists of corner-sharing CuO$_6$ octahedra and (IO$_3$)$^-$ trigonal pyramids (Figure 1a,c). The CuO$_6$ octahedra are tetragonally distorted owing to the first-order Jahn-Teller (JT) effect of the electronic state of Cu$^{2+}$ ($d^9$).[38, 39] The two axial Cu–O bonds are elongated with Cu–O bond lengths of 2.372(12) Å and 2.404(2) Å, while the four equatorial Cu–O bonds are compressed with the bond lengths ranging from 1.950(10) Å to 1.997(3) Å. Each I$^{5+}$ cation is bonded to three oxygen atoms in a trigonal pyramidal geometry induced by the second-order JT distortion, that is, the stereo-active lone-pair electrons of the I$^{5+}$ cation ($s^2$).[40, 41] The I–O bond lengths range from 1.804(11) Å to 1.838(12) Å. The lattice symmetry of α-Cu(IO$_3$)$_2$ is electronically governed by both the first and second-order Jahn-Teller distortions. As a result, α-Cu(IO$_3$)$_2$ exhibits a 2D electronic structure displaying strong overlap between Cu$^{2+}$-$d_{x^2-y^2}$, O-$p$, and I-$s$ orbitals (Figure 1e). This effective reduction in dimensionality between the 3D crystal structure and the 2D electronic structure is a consequence of decreased orbital degrees of freedom in Cu$^{2+}$.[22] The magnetic Cu$^{2+}$ cations also form triangular sublattice geometry, facilitating competing exchange interactions in this material (Figure 1e).

The crystal structure of Mn(IO$_3$)$_2$ can be described as a 3D framework formed by corner-sharing MnO$_6$ octahedra and (IO$_3$)$^-$ trigonal pyramidal units (Figure 1b,d). The unit cell of Mn(IO$_3$)$_2$ is comprised of two nonequivalent Mn$^{2+}$, with each Mn$^{2+}$ cation bonded to six oxygen in a slightly distorted octahedral geometry. The Mn-O bond lengths range from 2.112(2) Å to 2.213(3) Å. Each I$^{5+}$ ($s^2$) cation is bonded to three oxygen atoms in a trigonal pyramidal geometry driven by the second-order JT distortion,[40, 42] with I-O bond lengths ranging from 1.783(3) Å – 1.853(3) Å. The subtle distortion of MnO$_6$ octahedra is likely induced by the stereo-active lone-pair electrons of the I$^{5+}$ cation.[40] The crystal structure and electronic structure of Mn(IO$_3$)$_2$ are both 3-dimensional (Figure 1f). This is because in high-spin Mn$^{2+}$ $d^5$ system, the $d$ orbitals are evenly populated, resulting in zero orbital contribution. Consequently, the overlap between Mn$^{2+}$-$d$, O-$p$ and I-$s$ orbitals in Mn(IO$_3$)$_2$ is not directional as in α-Cu(IO$_3$)$_2$.

The macroscopic electric polarization of α-Cu(IO$_3$)$_2$ and Mn(IO$_3$)$_2$ is primarily produced from the alignment and addition of the (IO$_3$)$^-$ local dipole moment along the $b$ axis (Figure 1c-d). The local trigonal pyramidal symmetry of (IO$_3$)$^-$ in these materials is confirmed to be $C_{3v}$, $\Gamma_{vib} = 2A_1 + 2E$, as demonstrated by the four vibrational modes in the FTIR spectra (Figure S3). Cu–O and Mn–O vibrations are observed at 493 cm$^{-1}$ and 497 cm$^{-1}$, respectively.

**Magnetization.** To probe the magnetic behavior of α-Cu(IO$_3$)$_2$ and Mn(IO$_3$)$_2$, temperature-dependent magnetic susceptibility ($\chi$) data were measured from $T$ = 2 K – 300 K at $\mu_0H$ = 5 T and $\mu_0H$ = 0.05 T (Figure 2). α-Cu(IO$_3$)$_2$ and Mn(IO$_3$)$_2$ undergo antiferromagnetic ordering at $T_N$ = 8 K and $T_N$ = 6 K, respectively. To evaluate the spin-state of the Cu$^{2+}$ and Mn$^{2+}$ cations and the strength of magnetic interaction in these polar magnets, their magnetic susceptibility data at $T > T_N$ were analyzed by using the Curie-Weiss law (Equation 1):

$$\chi(T) = \frac{C}{T - \theta_{CW}} + \chi_0 \tag{1}$$

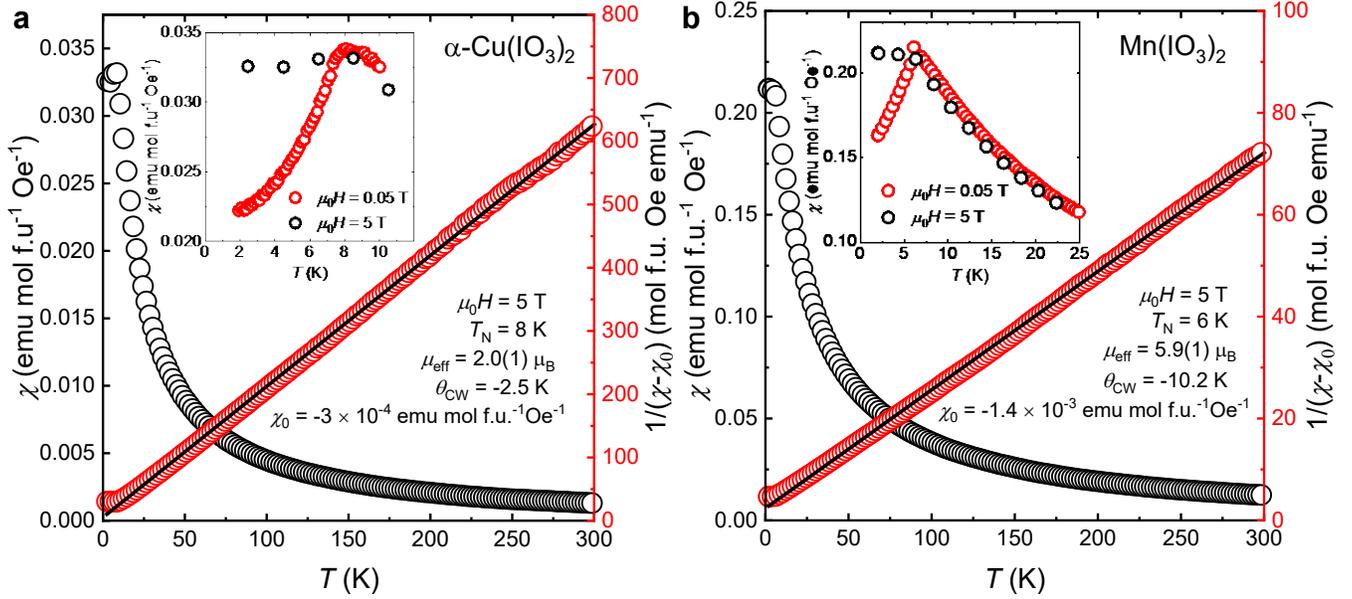

**Figure 2.** (black) Magnetic susceptibility of (a) α-Cu(IO$_3$)$_2$ and (b) Mn(IO$_3$)$_2$ measured from $T$ = 2 K – 300 K at $\mu_0 H$ = 5 T. (red) Curie-Weiss fitting of $1/(\chi-\chi_0)$ against temperature for the paramagnetic phase. The magnetic susceptibility measured at $\mu_0 H$ = 0.05 T is presented in the subset of the plot.

where $C$ is the Curie constant, $\theta_{CW}$ is the Curie-Weiss temperature and $\chi_0$ is the temperature-independent contribution to the magnetic susceptibility, which arise from small diamagnetic contribution of the core electrons and diamagnetic signals from the sample holder.[43, 44] The $\chi_0$ was estimated by adjusting it systematically until a linear plot of $1/(\chi-\chi_0)$ versus $T$ was obtained. The Curie-Weiss temperature $\theta_{CW}$ was extracted from the intercept of the linear fit to be -2.5 K for α-Cu(IO$_3$)$_2$ and -10.2 K for Mn(IO$_3$)$_2$. The negative values of the Curie-Weiss temperature indicate antiferromagnetic exchange interaction in both materials. The effective magnetic moment ($\mu_{eff}$) per M$^{2+}$ for α-Cu(IO$_3$)$_2$ and Mn(IO$_3$)$_2$ was estimated from Equation 2:

$$\mu_{eff} = \sqrt{\left(\frac{3k_B}{N_A}\right)C} \quad (2)$$

where $N_A$ is the Avogadro number, $k_B$ is the Boltzmann constant.[43] The effective magnetic moment obtained for Cu$^{2+}$ in α-Cu(IO$_3$)$_2$ is 2.0(1) $\mu_B$, which is higher than the ideal $g(S(S+1))^{1/2}$ = 1.73 $\mu_B$ value expected for a free $S$ = ½ moment but still within the limits of experimentally observed results for Cu$^{2+}$. This indicates the orbital angular momentum of Cu system is not completely quenched. Attempts were made to perform the fitting at two different temperature ranges 15 K ≤ $T$ ≤ 150 K and 150 K ≤ $T$ ≤ 300 K (Figure S8), resulting in the effective magnetic moments of 2.0(1) $\mu_B$ which are consistent with 2.0(1) $\mu_B$ obtained from the entire fitting range 8 K ≤ $T$ ≤ 300 K. For Mn(IO$_3$)$_2$, the effective magnetic moment obtained for Mn$^{2+}$ is 5.90(1) $\mu_B$, which is consistent with the expected magnetic moment for high spin $S$ = 5/2 (5.92 $\mu_B$), confirming no orbital contribution to the magnetic moment observed in this compound. These results of the magnetic behavior of α-Cu(IO$_3$)$_2$ and Mn(IO$_3$)$_2$ are strongly supported by their lattice symmetry and orbital overlap described in the structure discussion, highlighting the structure-property relationship of materials.

The upturn in temperature-dependent magnetic susceptibility ($\chi$) of α-Cu(IO$_3$)$_2$ and Mn(IO$_3$)$_2$ measured at high field $\mu_0 H$ = 5 T (Figure 2, Inset) suggests possible field-dependent magnetic behavior. To look into this further, field-dependent magnetization $M(H)$ data for α-Cu(IO$_3$)$_2$ and Mn(IO$_3$)$_2$ were measured. The hysteresis at $T < T_N$ = 8 K of the $M(H)$ curve of α-Cu(IO$_3$) indicates the possibility of competing AFM-FM interactions, potentially leading to a transition from the AFM state to a field-polarized magnetic state (Figure S9a).[25, 44, 45] The $M(H)$ curves of Mn(IO$_3$)$_2$ at $T < T_N$ = 6 K, however, show little, if any, hysteresis (Figure S9b).[44] While these deviating observations in α-Cu(IO$_3$)$_2$ and Mn(IO$_3$)$_2$ may be connected to the difference in spin and orbital aspects in these materials, this poses the query: How can the coupled spin and orbital degrees of freedom in these materials enhance or hinder the DM exchange interaction strength, and thus the presence or an absence of a transition from AFM order to a finite-field chiral spin state?

**Magnetoentropic Mapping.** To find answers to this question, we studied the magnetoentropic signatures of these complexes. The dc magnetization technique is applied, and data measured at different magnetic fields 0.4 T ≤ $\mu_0 H$ ≤ 4 T are depicted in Figure 3-4. The formation of field-induced topological magnetic spin textures typically corresponds to a magnetic anomaly in magnetoentropic curve attributed to first-order transitions to vortex-like spin states.[6, 32] These manifestations are observed as peaks or valleys near the magnetic transition temperature.[7, 32, 46] Magnetic phase diagram of a material can be mapped from the dc magnetization data by quantifying the entropy change associated with the magnetic transition under applied magnetic field.

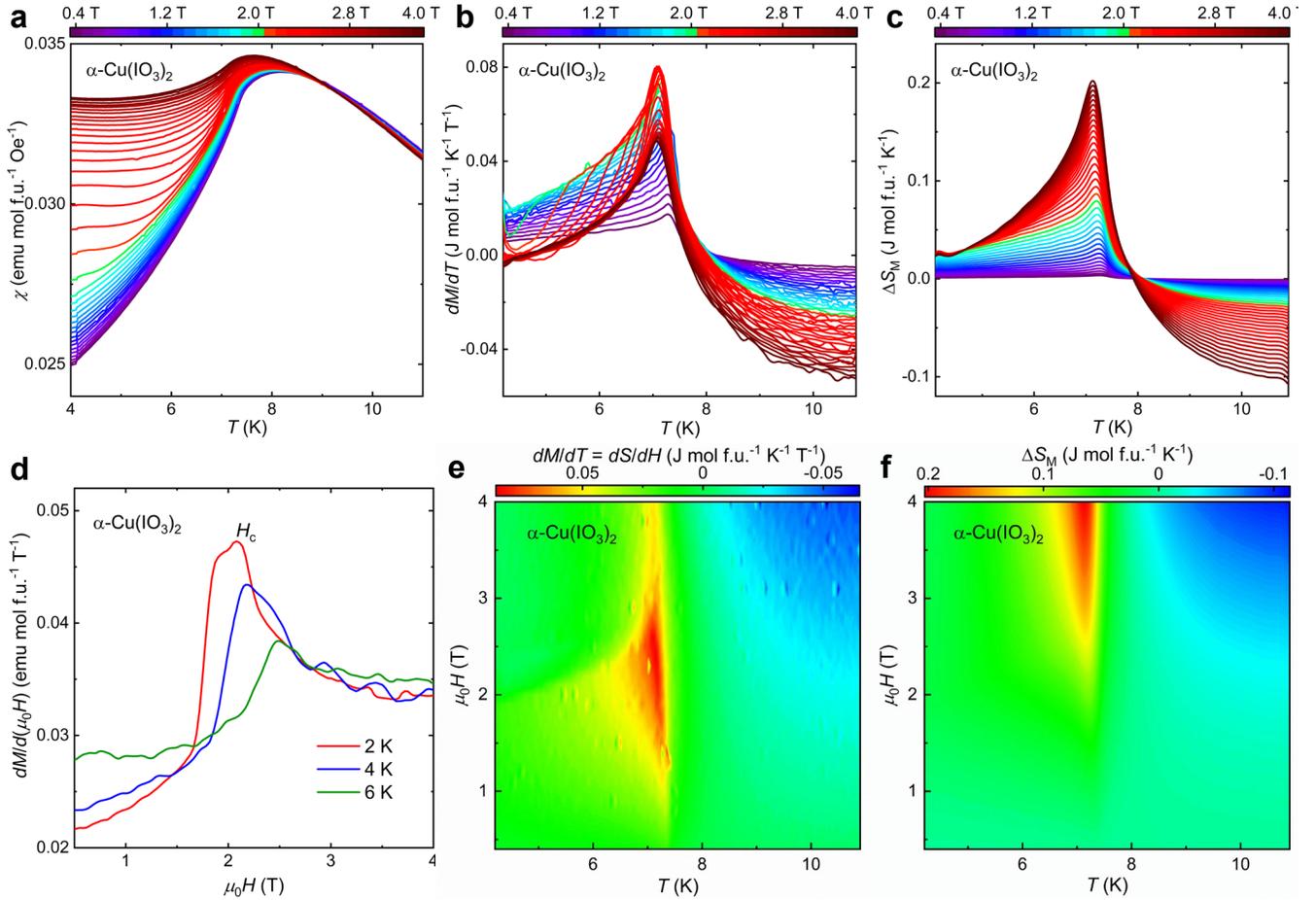

**Figure 3.** (a) $\chi(T)$ ($M(T)/(\mu_0 H)$) curves of α-Cu(IO$_3$)$_2$ at different magnetic field. (b) $dM/dT$ for α-Cu(IO$_3$)$_2$. (c) Isothermal magnetic entropy $\Delta S_M$ (H, T) of α-Cu(IO$_3$)$_2$ at each different magnetic field obtained by the integral of $dM/dT$ with respect to the magnetic field. (d) $dM/dH$ curve for α-Cu(IO$_3$)$_2$ showing critical fields $H_C$ at ~2 T – 3 T (e-f) Magnetoentopic map of α-Cu(IO$_3$)$_2$ near the ordering temperature. (e) A map of $dM/dT = dS/dH$ for α-Cu(IO$_3$)$_2$ reveals clear ridges (red) and valleys (blue) indicating first-order transitions. (f) A map of $\Delta S_M$ (H, T) of α-Cu(IO$_3$)$_2$ at $\mu_0 H$ = 0.4 T – 4 T.

The isothermal entropy change upon magnetization $\Delta S_M$ (H, T) is obtained from the Maxwell equation 3:[32]

$$\left(\frac{dS}{dH}\right)_T = \left(\frac{dM}{dT}\right)_H \quad (3)$$

where $S$ is the total entropy, $H$ is the magnetic field, $M$ is the magnetization and $T$ is the temperature. $dM/dT = dS/dH$ map provides a complementary elucidation to specific heat measurements for thermodynamic capacity. $\Delta S_M$ (H, T) is then derived using the equation 4:

$$\Delta S_M (H,T) = \int_0^H \left(\frac{dM}{dT}\right)_{H'} dH' \quad (4)$$

To extract magnetic entropy change, high density data of dc magnetization were collected on bulk samples of α-Cu(IO$_3$)$_2$ and Mn(IO$_3$)$_2$ from $T$ = 4 K to $T$ = 11 K using the VSM option of PPMS. For α-Cu(IO$_3$)$_2$, $\chi(T)$ curves under a series of applied magnetic fields are presented in Figure 3a. The upturn in the $\chi(T)$ curves at high field indicates evolution towards a field-polarized spin state in α-Cu(IO$_3$)$_2$. Further examination of the magnetic behavior of α-Cu(IO$_3$)$_2$ was performed by having a careful look at the $dM/dT$ (Figure 3b) and $\Delta S_M$ (Figure 3c) curves. Evident peaks are observed in these curves at 6.5 K ≤ $T$ ≤ 7.5 K and 2.0 T ≤ $\mu_0 H$ ≤ 3.5 T, and their sudden onsets imply first-order transitions. The magnetic anomalies in the $\Delta S_M$ curves of α-Cu(IO$_3$)$_2$ are comparable with those observed in Fe(IO$_3$)$_3$, MnGe and MnSc$_2$S$_4$.[26, 29, 32, 47, 48] Figure 3e depicts the $dM/dT = dS/dH$ map in which ridges and valleys in $dS/dH$ can imply field-induced first-order transitions and can assist in estimating entropy changes corresponding to these transitions. In the $\Delta S_M$ (H, T) map of α-Cu(IO$_3$)$_2$ (Figure 3f), the yellow virtually vertical line near $T$ ≈ 7.5 K suggests a boundary of first-order phase transitions between the partially spin-polarized state and the paramagnetic state. At $T$ < 7.5 K, a finite-field red region of positive entropy of approximately 0.2 J mol$^{-1}$ K$^{-1}$ is observed. The green region transitioning to blue at $T$ > $T_N$ = 8 K at higher fields corresponds to the paramagnetic state.

Let us briefly review connections between relevant topologically trivial vs. nontrivial spin states and entropy change in an effort to prepare for further discussion. While topologically trivial states such as helical and conical phases can form without an entropy change, the formation of topological distinct phases goes hand-in-hand with a change in entropy and a phase transition. However, it is worth keeping in mind that an entropy change is a necessary but not sufficient condition for realizing emergence of skyrmion phases.

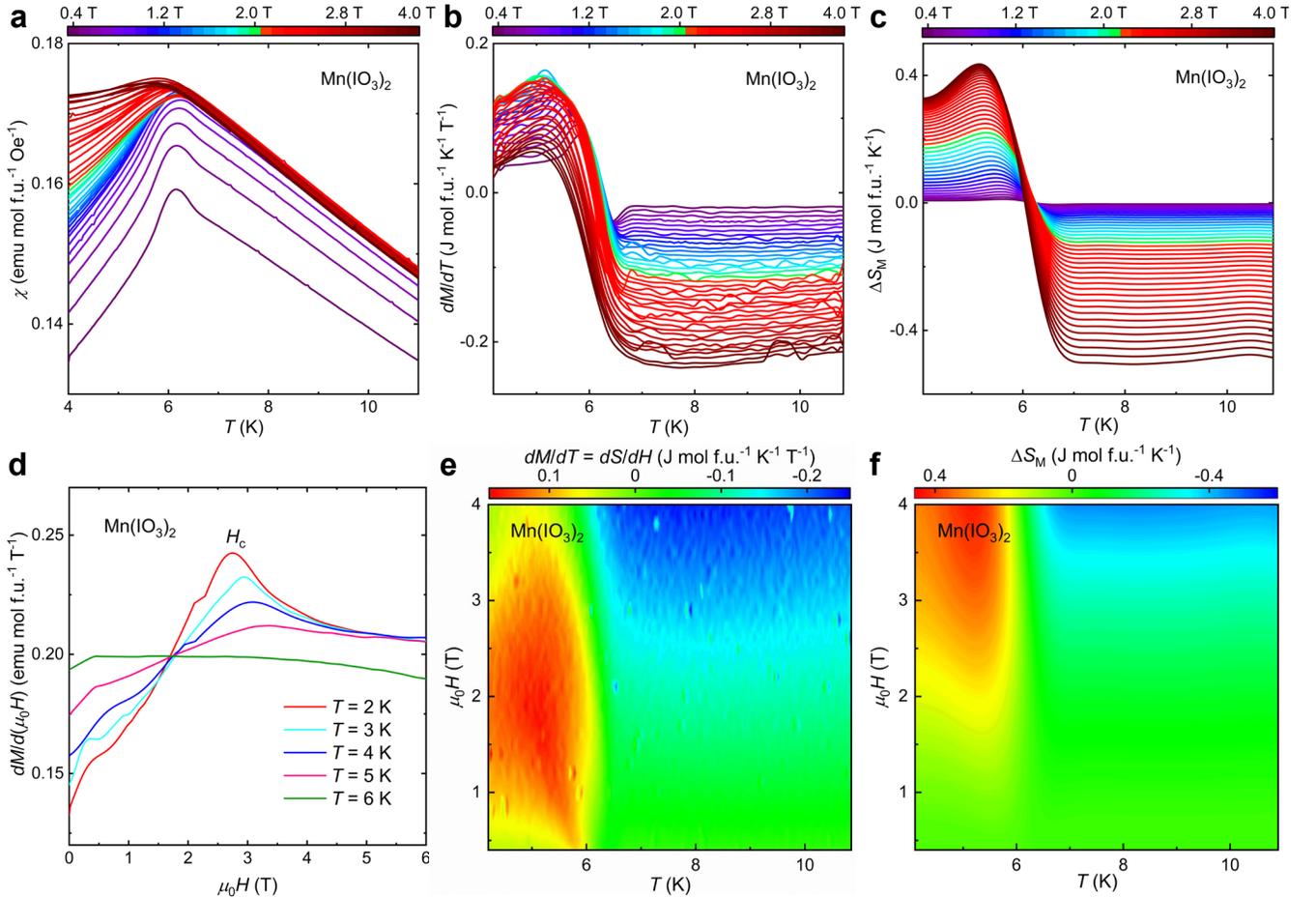

**Figure 4.** (a) $\chi(T)$ ($M(T)/(\mu_0 H)$) curves of Mn(IO$_3$)$_2$ at different magnetic field. (b) first derivative of magnetization with respect to the temperature, $dM/dT$, for Mn(IO$_3$)$_2$. (c) Isothermal magnetic entropy of Mn(IO$_3$)$_2$ at each different magnetic field obtained by the integral of $dM/dT$ with respect to the magnetic field. (d) $dM/dH$ curve for Mn(IO$_3$)$_2$ (e-f) Magnetoentopic map of Mn(IO$_3$)$_2$ near the ordering temperature. (e) A map of $dM/dT = dS/dH$ for Mn(IO$_3$)$_2$ reveals diffuse ridges (red) and valleys (blue) indicating competing AFM-FM exchange interaction. (f) A map of $\Delta S_M$ ($H$, $T$) of Mn(IO$_3$)$_2$ at 0.4 T ≤ $\mu_0 H$ ≤ 4 T.

It could be evolution of other topologically non-trivial spin textures such as meron and hopfion phases.[49, 50] While these results of the magnetoentropic mapping bear resemblance to evolution of finite-field skyrmion phases realized in other materials,[7, 17, 32, 51] the existence of vortex-like spin configurations in α-Cu(IO$_3$)$_2$ undoubtedly warrants additional study. The apparent finite-field first-order magnetic transitions in this material, nevertheless, can imply the enhanced DM exchange strength, which may tie in with allowed excitations between in the $^2$D ground state and excited states in α-Cu(IO$_3$)$_2$. This naturally invites the following thought: what if the spin multiplicity, degeneracy and symmetry of the ground state and first excited state in a system are modified, how does this influence evident anomalies which signify transitions to field-driven topological spin states? To answer, let us turn to Mn(IO$_3$)$_2$ which takes the states $^6$S in the ground state and $^4$G in the first excited state.

Figure 4a depicts how the magnetization of Mn(IO$_3$)$_2$ develops as a function of temperature under different fields near $T_N$ = 6 K. No apparent anomaly was observed in the $dM/dT$ and $\Delta S_M$ curves (Figure 4b-c), suggesting an ambiguous evidence of a first-order phase transition to a topologically nontrivial spin state. As shown in Figure 4d, the anomaly associated with the metamagnetic transition in Mn(IO$_3$)$_2$ is at $H_c$ = 0.5 and 3 T in the $dM/dH$ curve. In the $dM/dT = dS/dH$ and $\Delta S_M$ ($H$, $T$) maps of Mn(IO$_3$)$_2$ (Figure 4e-f), the region at $T > T_N$ = 6 K is green transitioning to blue is associated with the paramagnetic state. The diffused red region at $T < T_N$ = 6 K and 1.2 T ≤ $\mu_0 H$ ≤ 3.8 T can denote competing AFM-FM interactions.

Due to the paucity of strong evidence of a first-order phase transition, this appearance of increased entropy change is probably not associated with formation of vortex-like spin textures, but rather with competing AFM-FM exchange couplings. Although the possible evolution of a finite-field topological spin state is observed in α-Cu(IO$_3$)$_2$, a phase transition to field-driven winding spin configurations in Mn(IO$_3$)$_2$ is unclear. This can imply the role of the ground state and first excited state in enhancing the DM interaction, thereby potentially facilitating topologically protected spin formation.

**Specific Heat.** To investigate the thermodynamic properties of the ground state of α-Cu(IO$_3$)$_2$ and Mn(IO$_3$)$_2$, the zero field specific heat capacity was measured over the range of 2 K ≤ $T$ ≤ 300 K (Figure 5). An anomaly in the $C_p/T$ vs. $T$ plot for α-Cu(IO$_3$)$_2$ ($T_N$ ≈ 7.2 K) and Mn(IO$_3$)$_2$ ($T_N$ ≈ 5.0 K) is observed at approximately the temperature of the magnetic phase transition to long range AFM order of these materials as determined by magnetization and neutron diffraction experiments. This confirms that the transition observed in the $C_p/T$ vs. $T$ plot is magnetically driven.[29, 43, 52] The entropy recovered $\Delta S$ from the transition can be estimated from the equation 5:

$$\Delta S = \int_0^T \frac{C_V}{T} dT \qquad (5)$$

where $C_v$ is the heat capacity at constant volume, which is approximate to be $C_p$ – heat capacity at constant pressure – for solid at low temperatures and $T$ is the temperature. The phonon contribution to the magnetic specific heat and entropy can be estimated using an appropriate nonmagnetic isostructural material. For $Mn(IO_3)_2$, a nonmagnetic isostructural $Zn(IO_3)_2$ was used for phonon subtraction. In an attempt to evaluate how much entropy was recovered from the specific heat data at 2 K ≤ $T$ ≤ 300 K, $\Delta S_{mag}$ of $Mn(IO_3)_2$ was estimated to be 6.9(1) J mol f.u.$^{-1}$ K$^{-1}$, approximately 46 % of $S = 5/2$ value ($R\ln 6$)(Figure S10). The magnetic phase transition was not complete down to $T = 2$ K, hindering an accurate estimation of the entropy change $\Delta S_{mag}$ of $Mn(IO_3)_2$. For α-$Cu(IO_3)_2$, lack of appropriate nonmagnetic isostructural material, which could be used to account for phonon contribution, makes the accurate estimation of magnetic specific heat and entropy difficult.

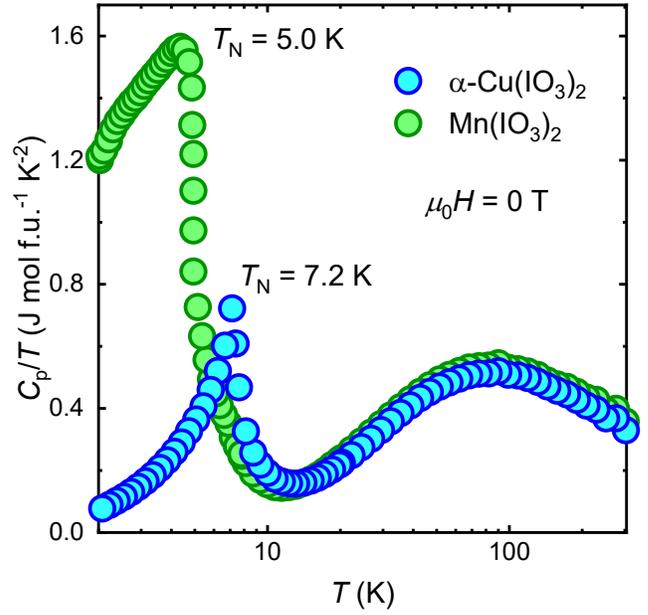

**Figure 5.** Molar heat capacity over temperature ($C_p/T$) vs. temperature ($T$) for α-$Cu(IO_3)_2$ and $Mn(IO_3)_2$.

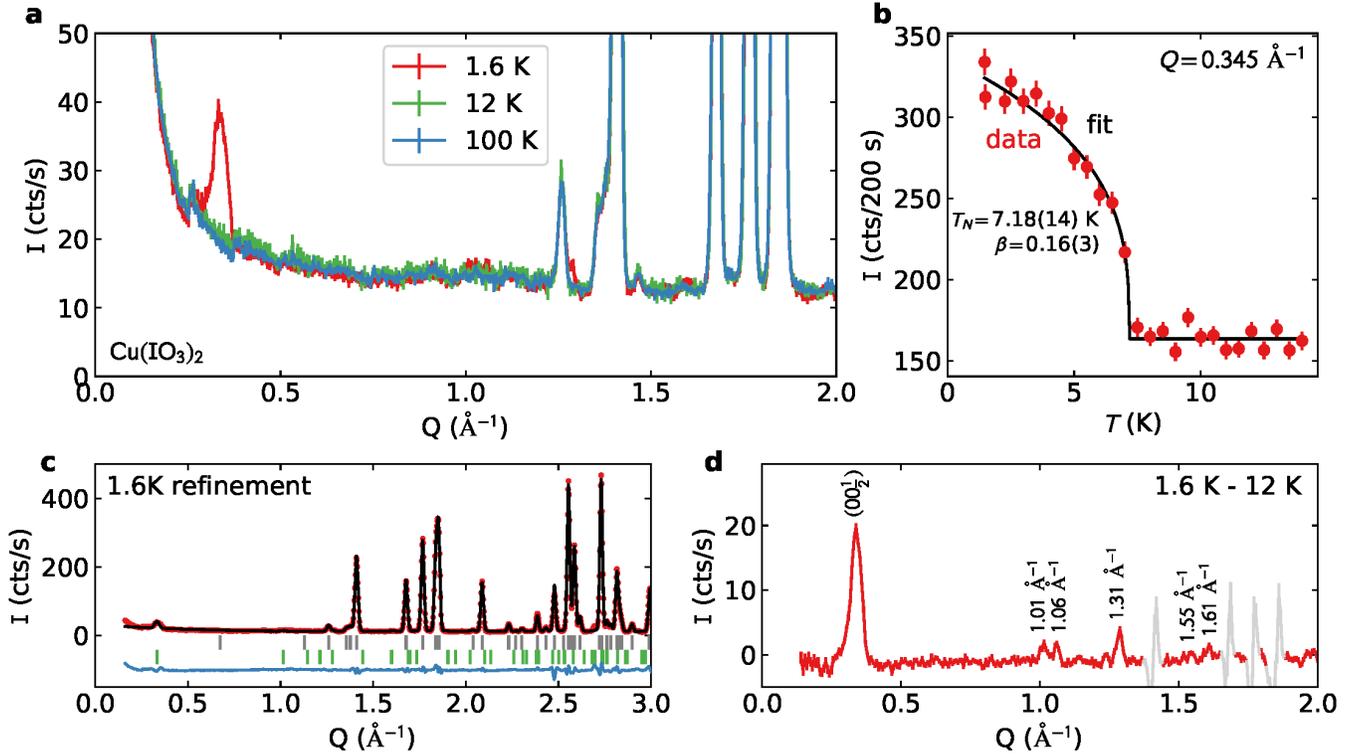

**Figure 6.** Neutron diffraction of α-$Cu(IO_3)_2$. Figure 6a shows the diffraction pattern above and below the Neel temperature, along with the diffraction at much larger temperatures. A strong magnetic Bragg peak appears at 0.345 Å$^{-1}$, but the other Bragg intensities are very weak. Figure 6b shows the order parameter curve of the 0.345 Å$^{-1}$ peak. The critical exponent is difficult to fit because of the few temperature points in the critical region, and thus is unreliable. 6c shows the refinement to the nuclear structure, assuming a commensurate k = (0,0,1/2) magnetic order. 6d shows the temperature-subtracted magnetic data, showing a family of incommensurate Bragg intensities. No single k-vector was found to match all of them.

**Neutron Diffraction.** In order to gain further insight into the magnetic order ground state of these materials, neutron diffraction experiments were performed on the HB2A diffractometer at Oak Ridge National Laboratory. By comparing the high and low temperature neutron diffraction patterns, it is evident that both α-Cu(IO$_3$)$_2$ and Mn(IO$_3$)$_2$ exhibit long range magnetic order at low temperatures. The α-Cu(IO$_3$)$_2$ magnetic structure is considerably more complex (Figure 6). One strong magnetic Bragg peak appears at $Q$ = 0.33(1) Å$^{-1}$, which matches antiferromagnetic k = (0,0,1/2) order. However, the rest of the magnetic Bragg peaks are small and are at non-integer wavevectors. The proximity to commensurate wavevectors is extremely close, which means this data indicates a long-wavelength incommensurate order, similar to Fe(IO$_3$)$_3$.[29] We performed an exhaustive search of all symmetry-allowed propagation vectors, but were unable to find a single propagation vector which matches all the observed magnetic peaks. We conclude that the ground state magnetism involves multiple axes of incommensurability in a multi-k magnetic structure. Because these incommensurate peaks are so weak, uniquely identifying the various propagation vectors will require a single-crystal diffraction experiment: too many possible multi-k structures exist for powder. If we ignore the incommensurability and fit the magnetism to a k = (0,0,1/2) structure using FullProf,[53] we find an ordered moment 0.69(2) μ$_B$ (Figure 6c). The true static ordered moment will be larger than this as this refinement neglects the some incommensurate magnetic intensity, but it is broadly consistent with a theoretical fully static moment $g(1/2)$ = 1.0 μ$_B$. In spite of the undetermined magnetic structure, it is clear that the α-Cu(IO$_3$)$_2$ magnetism is incommensurate with local antiferromagnetic magnetization near k = (0,0,1/2). This incommensurability can be attributed to the orbital contribution and anisotropic magnetic exchange of this material, which is supported by its aforementioned structure. The incommensurability of α-Cu(IO$_3$)$_2$ comports with the potential finite-field topological phase seen in Figure 3.

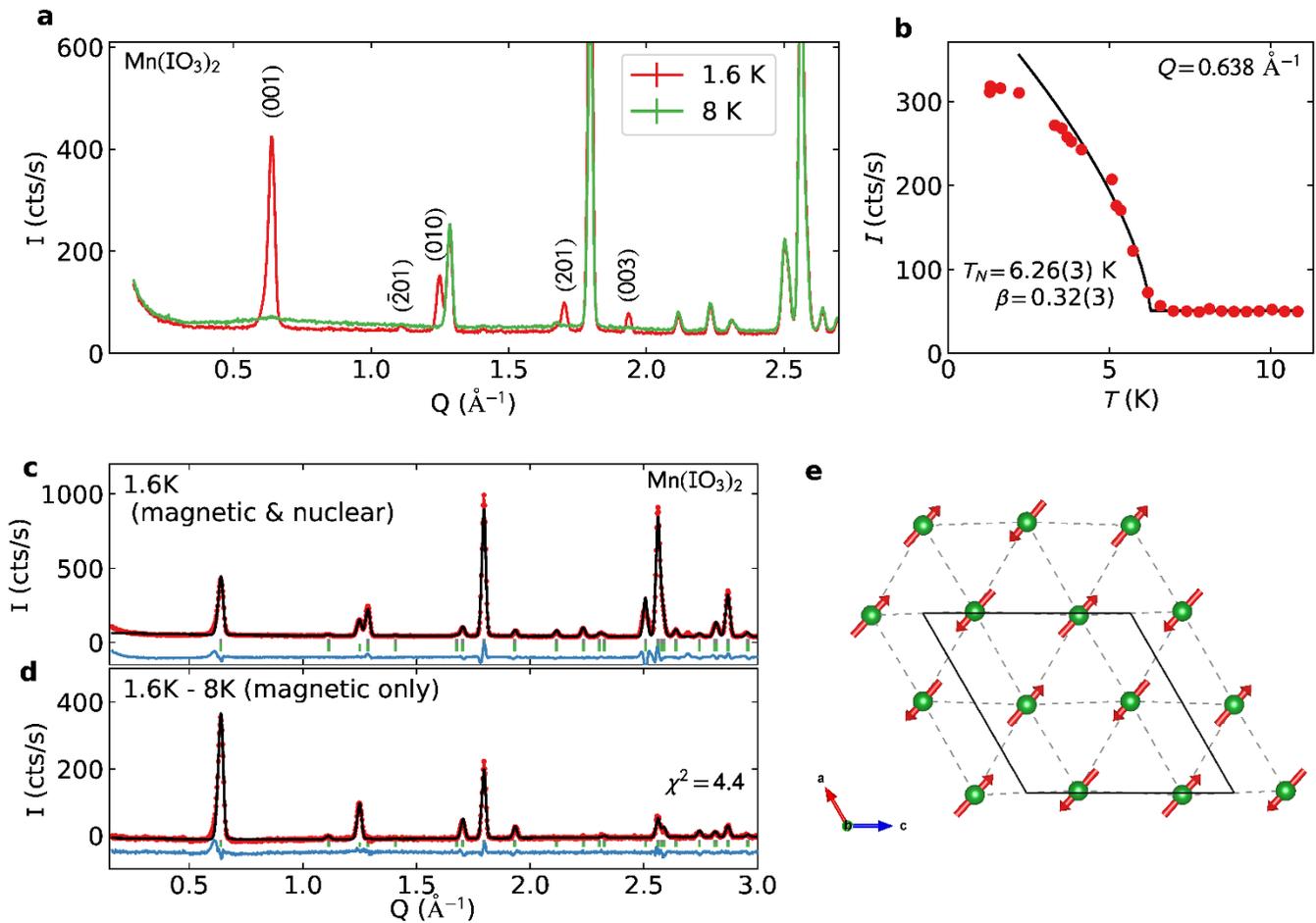

**Figure 7.** Neutron diffraction of Mn(IO$_3$)$_2$. Figure 7a shows the diffraction pattern above and below the Neel temperature, revealing new Bragg intensity with commensurate k = (0,0,0) order. 7b shows the order parameter curve of the (001) peak, which shows a critical exponent broadly consistent with 3D magnetic order. Figures 7c and 7d show the powder refinement of the nuclear and magnetic structures, showing a very good model fit to the magnetic Bragg intensities. Figure 7e shows the stripe antiferromagnetic ground state magnetic order.

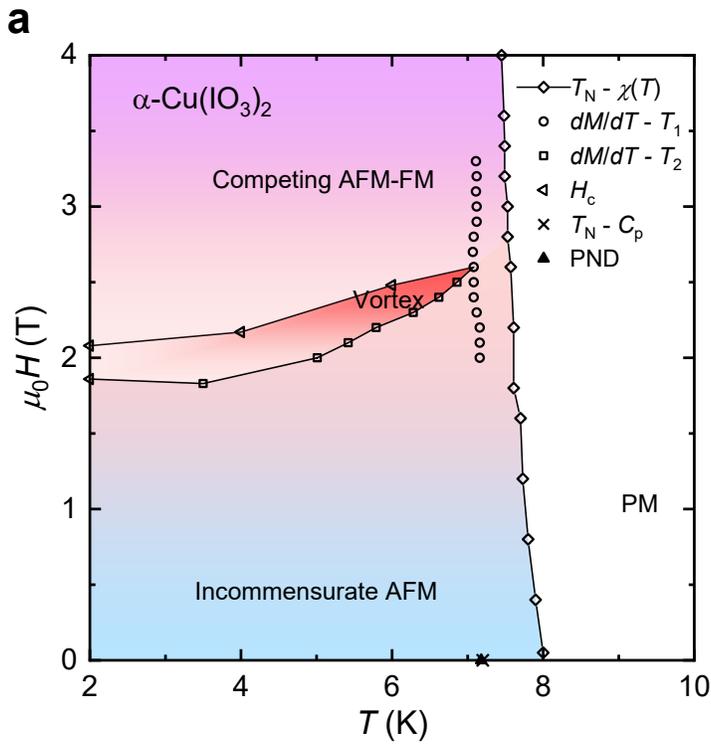
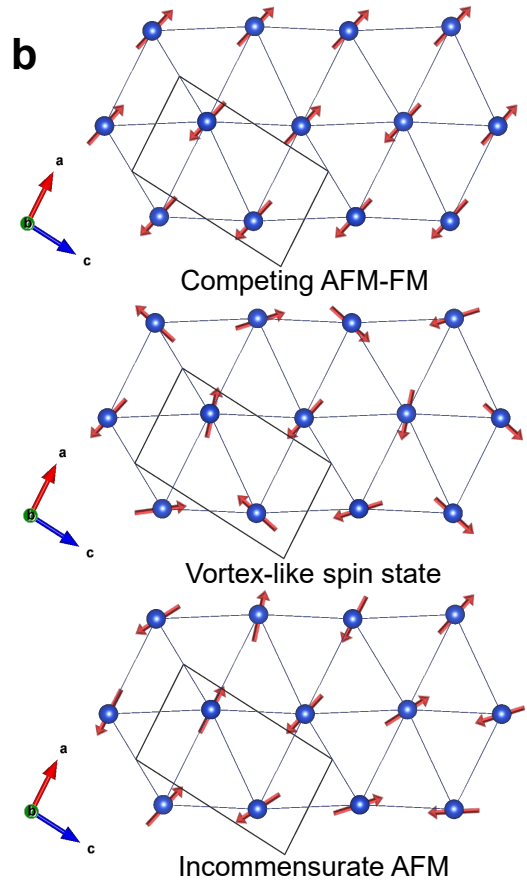
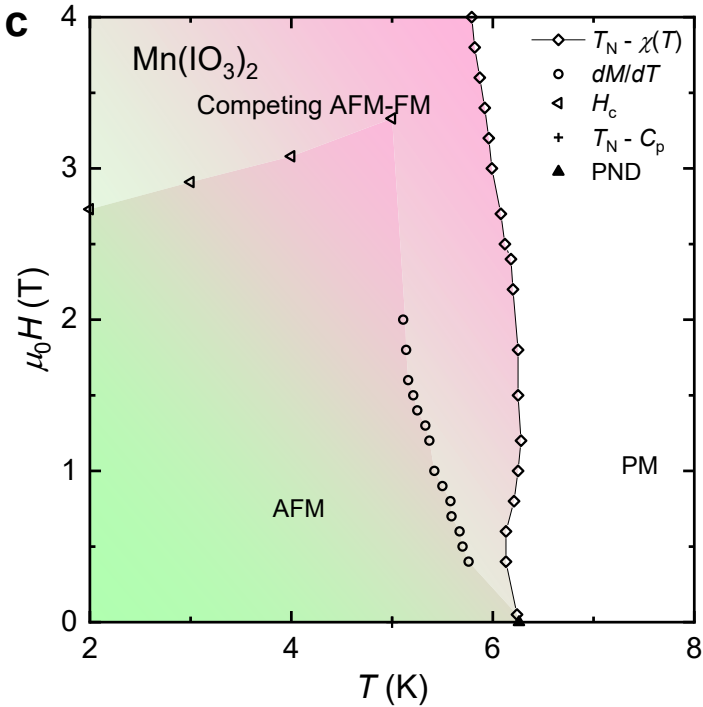
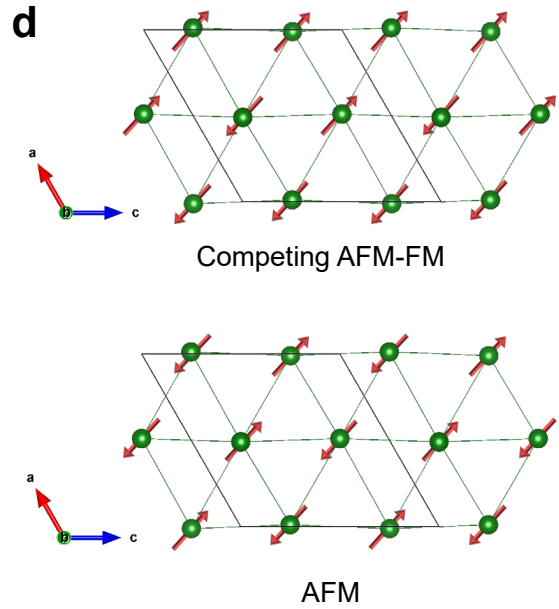

**Figure 8.** (a,c) Proposed magnetic phase diagram of (a) α-Cu(IO$_3$)$_2$ and (c) Mn(IO$_3$)$_2$ derived from the results of magnetic susceptibility, $dM/dT$, $dM/dH$, specific heat and neutron diffraction measurements. (b,d) Cartoon representation of the magnetic spin states of (b) α-Cu(IO$_3$)$_2$ and (d) Mn(IO$_3$)$_2$.

On the other hand, the Mn(IO$_3$)$_2$ magnetic order is the simpler of the two: comparisons of magnetic intensity above and below the Neel temperature shows new peaks appearing at integer wavevectors, indicating a propagation vector k = (0,0,0) (Figure 7). We decomposed the magnetic space group into irreducible representations using BasIrreps from the FullProf suite,[53] and found two irreps consistent with this propagation vector: ferromagnetic alignment of spins, and antiferromagnetic alignment. Using the FullProf suite to perform a magnetic Rietveld refinement, only the antiferromagnetic order matches the peak intensities observed, and it matches extremely well.

The resulting magnetic structure is shown in Figure 7e, with the refinement shown in Figure 7c and 7d. This commensurate magnetic order of Mn(IO$_3$)$_2$ is consistent with quenched orbital magnetism and isotropic magnetic exchange, which is in agreement with its electronic structure discussed earlier. The static magnetic moments refine to 4.26(17) $\mu_B$ and 4.47(18) $\mu_B$ for the two Mn sites, with stripe antiferromagnetic order primarily along the [101] direction. The magnetic moments are slightly reduced from the fully static $g(5/2) = 5.0$ $\mu_B$, which indicates some degree of fluctuating magnetism at 1.6 K. Nearly full static magnetic moments (~87%) for Mn(IO$_3$)$_2$ are observed in the neutron diffraction experiment at the base temperature T = 1.6 K, while only approximately half of them are obtained by the entropy chance extracted from the specific heat measurement down to T = 2 K. The variation between these results can be attributed to the following three reasons: (i) slight difference in the base temperatures at which the measurements were performed, (ii) the high resolution of neutron diffraction experiments and (iii) the imperfection of adiabaticity in specific heat experiments.

We proposed magnetic phase diagrams based on the results of magnetic susceptibility, $dM/dT$, $dM/dH$, specific heat and neutron diffraction measurements (Figure 8). For α-Cu(IO$_3$)$_2$, paramagnetic state is observed at $T > T_N = 8$ K and the ground state magnetic structure at zero field is incommensurate AFM as evidenced by the neutron diffraction experiment. Under finite fields, a possible emergence of a topological spin texture is predicted at 5.5 K ≤ $T$ ≤ 7.0 K and 2.0 T ≤ $\mu_0 H$ ≤ 2.5 T, whereas competing AFM-FM likely occurs at higher field $\mu_0 H$ > 2.5 T. Mn(IO$_3$)$_2$ displays a stripe AFM magnetic ground state at $T$ ≤ 6 K and $\mu_0 H$ = 0 T, a conclusion deduced from neutron experiment, and paramagnetic state at $T > T_N$. At high field $\mu_0 H$ > 3.0 T, competing AFM-FM magnetic state is proposed.

To provide a reasonable explanation for the difference in the magnetic phase diagrams of these materials, we take into account the ligand field theory (Figure 9). While d$^9$ single ion has $^2$D ground state ($S = ½$, $L = 2$), the Cu system with d$^9$ in D$_{4h}$ symmetry possesses no orbital degeneracy in the $^2$A$_{1g}$ ground state, thus the orbital angular momentum should be frozen. Although d$^5$ ion features zero orbital angular momentum in the ground state ($^6$S for single ion, and $^6$A$_{1g}$ for high spin d$^5$ in O$_h$ symmetry), orbital angular momentum in the first excited state ($^4$G) is non-zero. The ground state orbital magnetic moment is mostly quenched in α-Cu(IO$_3$)$_2$ and completely quenched in Mn(IO$_3$)$_2$ (the Curie-Weiss effective moment in the Cu material exceeds the spin-only value, suggesting a small orbital contribution to Cu magnesium), nevertheless, these systems are expected to have finite DM interaction attributed to virtual excitations to non-zero angular momentum states. Thus the departure in the $H$-$T$ magnetic phase diagrams of α-Cu(IO$_3$)$_2$ and Mn(IO$_3$)$_2$ is likely linked to the magnitude of the DM exchange, that is, the DM interaction of α-Cu(IO$_3$)$_2$ is more sizable than that of Mn(IO$_3$)$_2$ owing to spin-allowed excitations taking place in Cu system.

**Band Structure and Density of States.** To bolster connections between the aforementioned physical phenomena, crystal and magnetic structures and chemical bonding of these two polar magnets, DFT computations were performed using Wien2k[54] and the results are presented in Figure 10 and 11. Near the Fermi level ($E_F$ = 0 eV), the spin-polarized band structure of α-Cu(IO$_3$)$_2$ features some great dispersion, indicating the strong and directional orbital overlap of this material (Figure 10a-b). Mn(IO$_3$)$_2$, nevertheless, exhibits rather flat bands near $E_F$, supporting its 3D electronic structure and isotropic chemical bonds (Figure 11a-b). Figure 10c-e and 11c-e demonstrate the density of states (DOS) of α-Cu(IO$_3$)$_2$ and Mn(IO$_3$)$_2$, respectively.

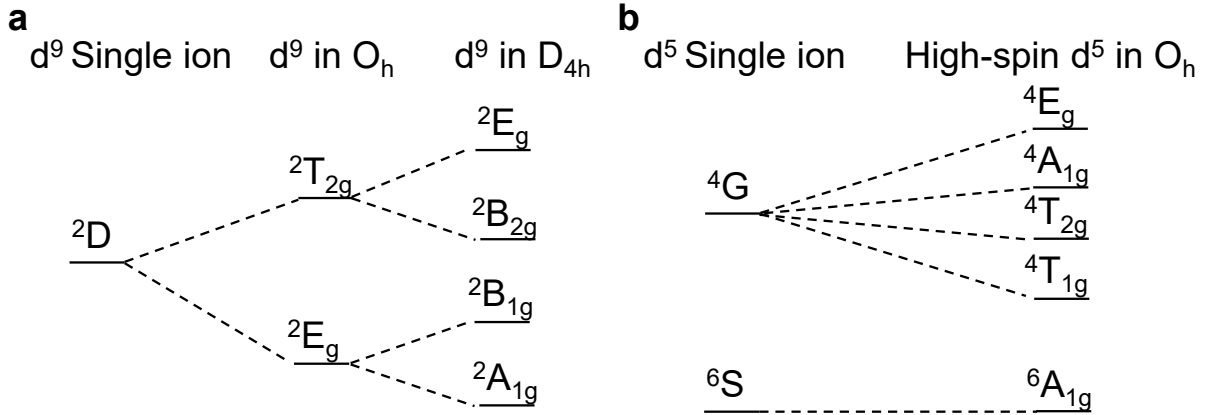

**Figure 9.** Energy states for (a) a d$^9$ ion in D$_{4h}$ symmetry and (b) a high-spin d$^5$ ion in O$_h$ symmetry based on the ligand field theory.

Overall, the DOS of Mn(IO$_3$)$_2$ are greater than that of α-Cu(IO$_3$)$_2$, originating from the narrower bandwidth observed in the Mn material. Cu(Mn)-*d* derived states are polarized as depicted in Figure 10c and 11c. The spin of these transition metals then polarizes the O-*p*, I-*s* and I-*p* electrons, resulting in Cu(Mn) network with AFM interactions. The valence band maximum is mainly composed of the Cu(Mn)-*d*, O-*p*, I-*s* and I-*p* states (Figure 10c-e, 11c-e). The results of DFT computations solidify the links between the magnetic exchange interactions as well as the structural and electronic structure of these materials as discussed above. While α-Cu(IO$_3$)$_2$ possesses highly anisotropic magnetic couplings attributed to its reduction in electronic dimensionality, Mn(IO$_3$)$_2$ displays isotropic spin communication network owing to its 3D electronic structure.

Anisotropic spin Hamiltonians, which give rise to an anisotropic exchange interaction including the asymmetric DM type, play a critical role in the observed high magnetic entropy in the *H-T* phase diagrams of these systems. There are two anisotropy energies: out-of-plane anisotropy and in-plane anisotropy. An out-of-plane anisotropy energy drives the spin to lie in the basal plane, an in-plane anisotropy energy, on a contrary, selects the orientation of the spins within the basal plane. These two anisotropic energies are expected to be higher in α-Cu(IO$_3$)$_2$ than those in Mn(IO$_3$)$_2$, giving rise to highly asymmetric magnetic exchange interactions in Cu system. This is supported by the greater dispersion in the spin-polarized band structure of α-Cu(IO$_3$)$_2$ compared to that of Mn(IO$_3$)$_2$.

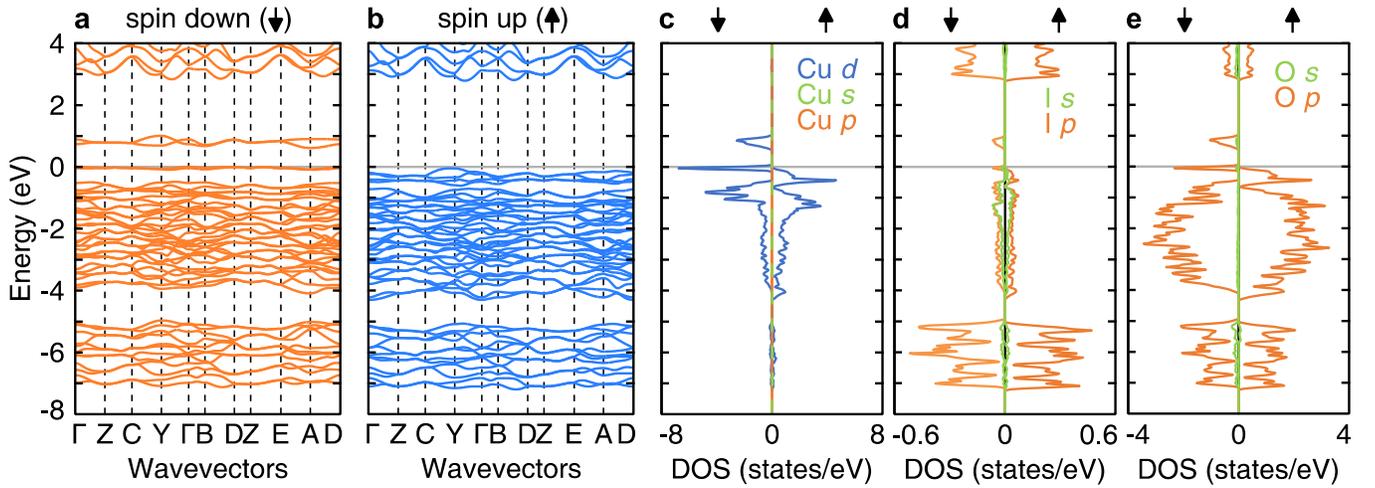

**Figure 10.** (a-b) Spin-polarized band structure and (c-e) density of states (DOS) of α-Cu(IO$_3$)$_2$ showing the orbital overlap of Cu-*d*, I-*s*, I-*p*, and O-*p*, states at the valence band maximum. The Fermi level is set at 0 eV.

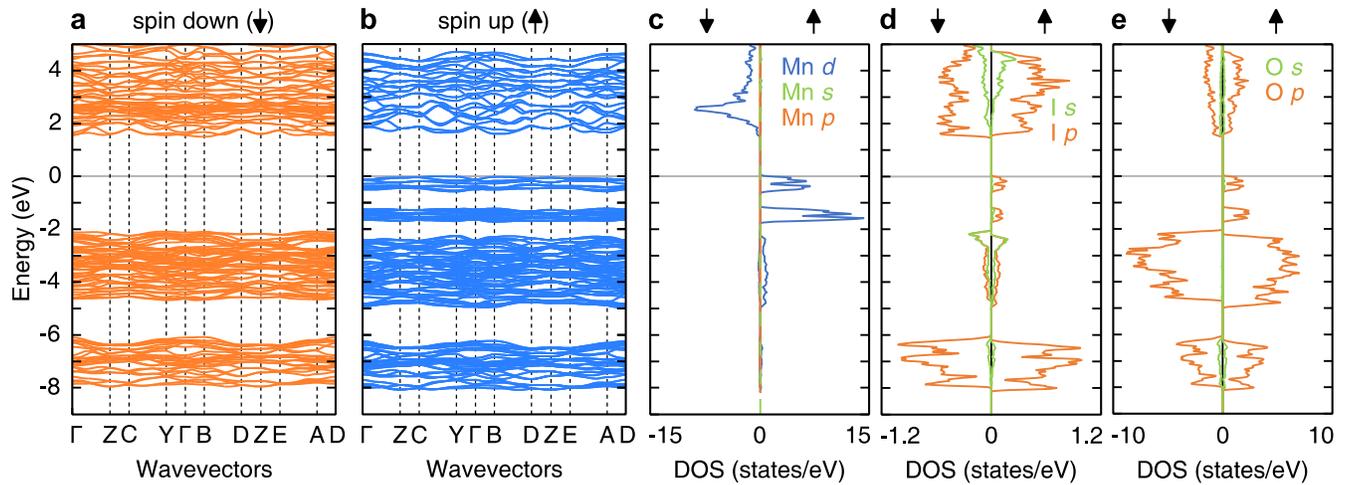

**Figure 11.** (a-b) Spin-polarized band structure and (c-e) density of states (DOS) of Mn(IO$_3$)$_2$ showing the orbital overlap of Mn-*d*, I-*s*, I-*p*, and O-*p*, states at the valence band maximum. The Fermi level is set at 0 eV.

Taken together, the overall picture highlights a vital position of the coupled spin and orbital aspects in the ground state and excited states in fine-tuning magnetic interaction to discover new topologically nontrivial spin phases. Plausible explanations for the dissimilarity in the *H-T* magnetic phase diagrams of Cu and Mn system can be multidimensional and are proposed as follows. First, it may be associated with the size of the DM exchange, that is, the DM interaction of α-Cu(IO$_3$)$_2$ is more appreciable that that of Mn(IO$_3$)$_2$ attributed to spin-allowed excitations occurring in α-Cu(IO$_3$)$_2$. Second, anisotropy of magnetic exchange interactions also likely contributes to incommensurability. The DFT computation work resulted in an evident discrepancy between the spin-polarized band structure and density of states of these two materials. α-Cu(IO$_3$)$_2$ displays strong and directional orbital overlap, whereas Mn(IO$_3$)$_2$ features 3D electronic structure and isotropic chemical bonding. Third, quantum vs classical spins can factor in the divergence in their magnetic properties. Small spin magnetic moment S = ½ in α-Cu(IO$_3$)$_2$ retains more quantum mechanical behavior, however larger spin number S = 5/2 in Mn(IO$_3$)$_2$ behaves more classically.

## CONCLUSION

Understanding how coupled degrees of freedom in polar magnets can enhance or suppress asymmetric exchange interactions, and consequently possible topological spin physics, has been the focus of considerable efforts. Our insights into this mechanism presented here make material progress in this understanding. Polar magnets, α-Cu(IO$_3$)$_2$ ($^2$D) and Mn(IO$_3$)$_2$ ($^6$S), were chosen for this study because of their variation in combined spin and orbital components in the ground state and excited states. The low-temperature preparation of pure phases of these two complexes assisted by the knowledge of their thermal behavior demonstrates a feasible method for synthesizing new polar magnetic materials. The results indicate that apparent phase transitions to possible topologically distinct spin states were observed in α-Cu(IO$_3$)$_2$ but unclear in Mn(IO$_3$)$_2$. This is supported by their magnetic ground state deduced from neutron diffraction experiments. The magnetic structure of α-Cu(IO$_3$)$_2$ features a rather complex incommensurability, whereas Mn(IO$_3$)$_2$ exhibits a commensurate stripe AFM ground state. This can be attributed to multifaceted explanations including (i) more appreciable DM interaction in α-Cu(IO$_3$)$_2$ compared to that in Mn(IO$_3$)$_2$, (ii) highly anisotropic exchange coupling vs isotropic interaction and (iii) quantum vs classical spins. The effects of these coupled degrees of freedom in asymmetric exchange are mediated through the overlapping of the transition metal Cu-*d* or Mn-*d*, O-*p*, I-*s* and I-*p* states and the low dimensionality of α-Cu(IO$_3$)$_2$ electronic structure. This work provides a useful avenue for modifying the magnetic interaction at the atomic level, broadening our horizon for new knowledge as well as realization of novel states of matter.

## AUTHOR INFORMATION

### Corresponding Author
* Thao T. Tran, thao@clemson.edu
Department of Chemistry, Clemson University, Clemson, SC 29634, USA

### Author Contributions
The manuscript was written through contributions of all authors. All authors have given approval to the final version of the manuscript.

### Notes
The authors declare no competing financial interest.

## ACKNOWLEDGMENT

This work was supported by Clemson University, College of Science, Department of Chemistry. E.E.O. acknowledges the COSSAB-GIAR Grant from College of Science, Clemson University. T.T.T. thanks the 2021 Support for Early Exploration and Development (SEED) Grant. Research at Gdansk University of Technology was supported by the National Science Center (Poland) under SONATA-15 grant (no. 2019/35/D/ST5/03769). MJW gratefully acknowledges the Ministry of Science and Higher Education scholarship for young scientists. A portion of this research used resources at the High Flux Isotope Reactor, a DOE Office of Science User Facility operated by the Oak Ridge National Laboratory. Use of the Advanced Photon Source at Argonne National Laboratory was supported by the U. S. Department of Energy, Office of Science, Office of Basic Energy Sciences, under Contract No. DE-AC02-06CH11357. This manuscript has been authored by UT-Batelle, LLC, under contract DE-AC05-00OR22725 with the US Department of Energy (DOE). The US government retains and the publisher, by accepting the article for publication, acknowledges that the US government retains a nonexclusive, paid-up, irrevocable, worldwide license to publish or reproduce the published form of this manuscript, or allow others to do so, for US government purposes. DOE will provide public access to these results of federally sponsored research in accordance with the DOE Public Access Plan (http://energy.gov/downloads/doe-public-access-plan).

## ABBREVIATIONS

PXRD, Powder X-ray diffraction, ATR-FTIR, Attenuated total reflection Fourier transform infrared, PPMS, Physical Properties Measurement System; TGA, Thermogravimetric analysis; DSC, Differential scanning calorimetry; SOC, Spin-orbit coupling.

Table of Content

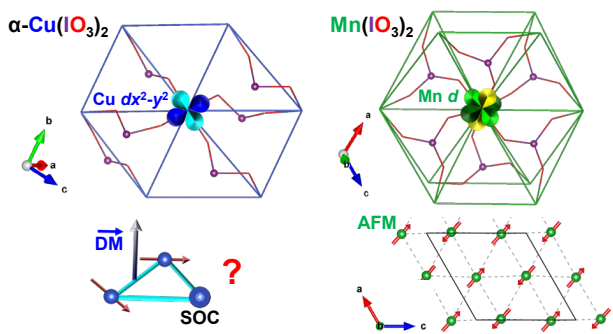

The effects of spin and orbital components to enhance asymmetric exchange interactions in polar magnets, α-Cu(IO$_3$)$_2$ ($^2$D) and Mn(IO$_3$)$_2$ ($^6$S) are reported. The magnetic structure of α-Cu(IO$_3$)$_2$ features a long-wavelength incommensurability, whereas Mn(IO$_3$)$_2$ exhibits a commensurate stripe AFM ground state. The work demonstrates connections between combined spin and orbital degrees of freedom, magnetic coupling dimensionality and asymmetric exchange, providing a useful approach for tuning asymmetric interaction which promotes evolution of topologically distinct spin phases.